\documentclass[twocolumn,bibnotes,superscriptaddress,showpacs,amsmath,amssymb,aps,prl]{revtex4-2}

\usepackage{graphicx}
\usepackage{amsmath}
\usepackage{amssymb}
\usepackage{color}
\usepackage{comment}
\usepackage[normalem]{ulem}
\usepackage{bm}

\newcommand{\zf}{$Z_2$ flux}

\newcommand{\ru}{$\alpha$-$\rm{RuCl}_3$}
\newcommand{\na}{$\textrm{Na}_2\textrm{Co}_2\textrm{Te}\textrm{O}_6$}

\newcommand{\ha}{$\bm{H}\parallel \bm{a}$}
\newcommand{\has}{$\bm{H}\parallel \bm{a^{\ast}}$}
\newcommand{\co}{Co$^{2+}$}

\usepackage[colorlinks=true,linkcolor=blue,citecolor=blue]{hyperref}

\begin{document}

\title{Field-Angle-Resolved Specific Heat in Na$_2$Co$_2$TeO$_6$: Evidence against Kitaev Quantum Spin Liquid}

\author{Shengjie~Fang}\email{fang@qpm.k.u-tokyo.ac.jp}
\author{Kumpei~Imamura}
\affiliation{Department of Advanced Materials Science, University of Tokyo, Kashiwa, Chiba 277-8561, Japan}

\author{Yuta~Mizukami}
\affiliation{Department of Advanced Materials Science, University of Tokyo, Kashiwa, Chiba 277-8561, Japan}
\affiliation{Department of Physics, Tohoku University, Sendai 980-8578, Japan}

\author{Ryuichi~Namba}
\author{Kota~Ishihara}
\author{Kenichiro~Hashimoto}
\author{Takasada~Shibauchi}\email{shibauchi@k.u-tokyo.ac.jp}
\affiliation{Department of Advanced Materials Science, University of Tokyo, Kashiwa, Chiba 277-8561, Japan}

\begin{abstract}
Kitaev quantum spin liquids (KSLs) in layered honeycomb magnets are known to host Majorana quasiparticles, whose excitations depend strongly on the direction of the applied magnetic field. In the high-field phase of $\alpha$-RuCl$_3$, specific heat measurements have revealed characteristic field-angle dependence of low-energy excitations consistent with the Kitaev model, providing bulk evidence for the KSL state. Here we present low-temperature measurements of specific heat $C(T)$ for another KSL candidate \na\ (NCTO) under field rotation within the honeycomb plane. Above the critical field of antiferromagnetic order, the field-angle dependence of $C/T$ exhibits minima along the bond directions, contrasting with the maxima observed in the KSL state of $\alpha$-RuCl$_3$. Our analysis indicates nodeless, fully-gapped excitations, which are inconsistent with the angle-dependent Majorana excitations with gapless nodes predicted by the Kitaev model. These findings suggest that low-energy excitations in NCTO are governed by gapped magnon excitations rather than Majorana quasiparticles, providing thermodynamic evidence against a KSL state. 
\end{abstract}

\maketitle

In quantum spin liquids, no long-range magnetic order appears down to absolute zero, and the ground states have many-body entanglement which can lead to exotic excitations \cite{Anderson1973,Savary2017}. These phenomena result from quantum fluctuations and frustration inherent to the lattice structures, and ground states in these systems are usually difficult to solve. However, the Kitaev model with two-dimensional honeycomb structures with bond-dependent Ising interactions offers exactly solvable quantum spin liquid ground states by mapping spin-1/2 operators to Majorana fermions that are their own antiparticles\,\cite{Kitaev2006}. Within this framework, two types of Majorana fermions emerge: one is localized on a honeycomb lattice, known as \zf\ (vison), and the other is the itinerant Majorana quasiparticle\,\cite{Motome2020}. The coexistence of both \zf\ and itinerant Majorana fermions on the honeycomb lattice culminates in the emergence of non-Abelian anyons, which are anticipated to serve as building blocks for fault-tolerant topological quantum computing. Therefore, the realization of the KSLs in real materials is one of the major issues in condensed matter physics.

At present, the most prominent candidate material for realizing a KSL is the honeycomb magnet \ru, which is a spin-orbit assisted Mott insulator with $4d$ electrons \cite{Do2017,Kubota2015a,Majumder2015,Winter2016,Kim2016,Ran2017,Banerjee2018}. Experimental research in the high-field state of \ru\ has reported the half-integer quantized thermal Hall effect \cite{Kasahara2018,Yamashita2020,Yokoi2021,Bruin2022}, which is a signature of chiral edge mode of Majorana fermions expected in the KSL. However, the thermal Hall conductivity $\kappa_{xy}$ in \ru\ turned out to have some sample dependence, which leads to strong debates on the origins of $\kappa_{xy}$ \cite{Lefrancois2022,Czajka2023,Kasahara2022}. 

Recently, a different approach has been developed to study the KSL focusing on the characteristic field-angle dependence of the low-energy excitations of itinerant Majorana quasiparticles \cite{Tanaka2022,Hwang2022,Imamura2024}. In the Kitaev model, the itinerant Majorana quasiparticles exhibit Dirac-cone type gapless dispersions at zero field, but the application of magnetic field opens the excitation gap $\Delta_{\mathrm{M}}$. This Majorana gap $\Delta_{\mathrm{M}}$ is given by 
\begin{equation}
    \Delta_{\mathrm{M}} \propto \frac{|h_x h_y h_z|}{\Delta ^2 _{\mathrm{flux}}},
    \label{deltam}
  \end{equation}
where $\Delta_{\mathrm{flux}} ( >\Delta_{\mathrm{M}})$ is the excitation gap of $Z_2$ fluxes, $h_x$, $h_y$, and $h_z$ denote the $x$, $y$, and $z$ components of the applied magnetic field on the spin axes, respectively. This immediately indicates that the Majorana gap becomes zero when the field direction is perpendicular to one of the spin axes ($h_x h_y h_z=0$). In the high-field state of \ru, high-resolution specific heat measurements under field rotation within the honeycomb plane ((111) plane on the spin-axis coordinate) have revealed the excitation gap following $|\cos(3\phi)|$ angle dependence (where $\phi=0$ corresponds to the field direction parallel to the zigzag direction) \cite{Tanaka2022}, with the gap closing behavior when the field is along the Ru-Ru bond direction ([-1,1,0]) \cite{Imamura2024}. This is fully consistent with the field-angle dependence of Majorana gap in Eq.\,(\ref{deltam}), providing strong evidence for the KSL from the bulk measurements. 

\begin{figure*}[!t]
    \includegraphics[width=\linewidth]{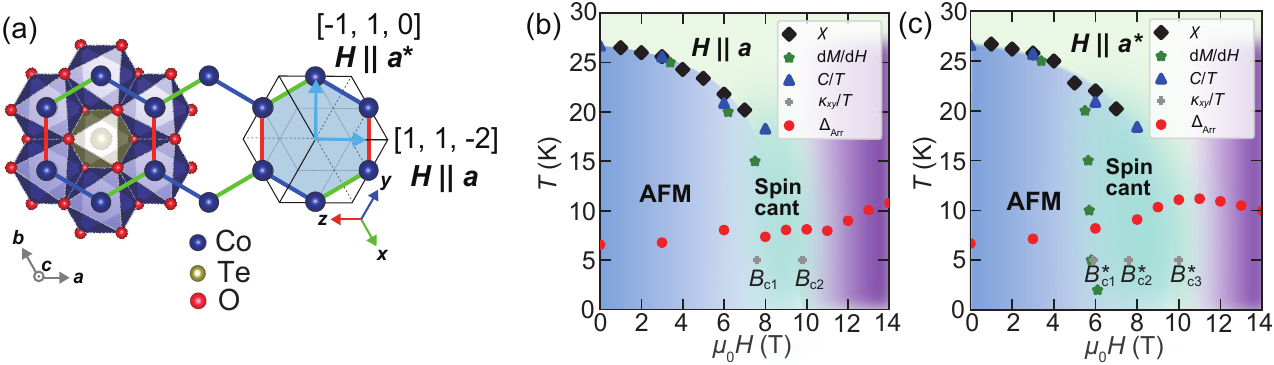}
    \caption{(a) Schematic crystal structure of \na\ with definitions of the crystallographic axes $(a, b, c)$ and the spin axes $(x, y, z)$.  
    (b,c) Temperature-field phase diagrams of \na\ for \ha\ (b) and \has ($\perp \bm{a}$) (c). The phase boundaries are determined by the peaks in the temperature dependence of magnetic susceptibility $\chi(T)$ (black diamonds), the field derivative of magnetization $dM/dH$ (green stars), and specific heat divided by temperature $C/T$ (blue triangles). We also plot the critical fields reported in the thermal Hall conductivity measurements (gray crosses) \cite{Takeda2022}. The red circles represent the low-energy excitation gap obtained from the analysis in Fig.\,\ref{Fig3}.}
    \label{Fig1}
\end{figure*}

Here we employ this approach to a different candidate material of KSLs, which is the Co-based layered honeycomb magnet \na\ (NCTO). In this $3d$ material, the magnetic ion of \co\ is surrounded by O$^{2-}$ ions and forms edge-sharing octahedra honeycomb layers \cite{Jackeli2009,Viciu2007}, as depicted in Fig.\,\ref{Fig1}(a). Recent theoretical calculations have shown that the electron hopping for the $e_{g}$ channel, which exists in the high-spin configuration of $3d^7$, suppresses the Heisenberg interaction $J$ and strengthens the Kitaev interaction $K$ \cite{Liu2018,Motome_2020,Sano2018}. This expectation has generated considerable interest in the magnetic properties of NCTO as a potential KSL candidate. We investigate the low-temperature specific heat $C$ under rotated magnetic fields within the honeycomb plane, from which the excitation gap is extracted as a function of field angle. Contrary to the expectation of the angle dependence of the gap in the KSL, we observe no gapless behavior for the field along the Co-Co bond direction (\has) in the high-field state of NCTO. The angular dependence of $C/T$ shows a maximum along the zigzag direction (\ha), an opposite behavior from the results in the high-field KSL phase of \ru. These results indicate that low-energy excitations in the high-field state of NCTO cannot be described by Majorana quasiparticles in the KSL.

As shown in the phase diagrams of NCTO in Figs.\,\ref{Fig1}(b) and (c), at zero field the antiferromagnetic (AFM) order sets in below the Néel temperature $T_{\rm N}=27$\,K \cite{Yao2020,Lin2021,Hong2021,Yao2020,Takeda2022,Bera2023}. The magnetic structure of the AFM phase has been studied by neutron scattering experiments, and a collinear single-$\bm{q}$ zigzag state\,\cite{Lefrancois2016,Bera2017,Yao2020} or non-collinear triple-$\bm{q}$ state\,\cite{Chen2021,Lee2021,Yao2023,Kruger2023} have been discussed. In any case, the AFM ground state at zero field indicates the existence of non-Kitaev terms in NCTO. It has also been discussed that the Kitaev interaction $K$ is significant in NCTO, although conflicting ferromagnetic ($K=-9.0$\,meV \cite{Songvilay2020}) and antiferromagnetic ($+3.6$\,meV \cite{Kim2021}) estimates have been reported. This AFM order can be suppressed by applying magnetic fields either along the $\bm{a}$ or $\bm{a^\ast}$ axis within the honeycomb plane, which leads to a canted AFM state and a possible spin-liquid phase \cite{Yao2020,Lin2021,Hong2021,Takeda2022,Bera2023}. From the kinks observed in the $\kappa_{xy}$ measurements, several critical fields between $\sim 6$ and $\sim10$\,T have been suggested \cite{Takeda2022}, which are also shown in Figs.\,\ref{Fig1}(b) and (c). Whether a KSL exists above $\sim10$\,T (up to the forced polarized phase above $\sim 14$\,T) has attracted much attention, which we aim to address through the measurements of field-angle dependence of specific heat. Note that the non-Kitaev terms can give an additional term $\propto |h_x+h_y+h_z|$ in Eq.\,(\ref{deltam}), which is however zero within the honeycomb (111) plane \cite{Tanaka2022,Hwang2022}. Thus, the in-plane field angle measurements of excitation gap can provide a stringent test of the KSL in honeycomb magnets.

High-quality NCTO single crystals were synthesized via the self-flux method (2D Semiconductors Inc.). The specific heat capacity $C$ of NCTO was measured using the long-relaxation method with the capability of field rotation by using a piezo-based rotator within a 14 T superconducting magnet \cite{Tanaka2022,Imamura2024}. The phonon contribution $\beta T^3$ at low temperatures has been estimated from the results in the isomorphic non-magnetic insulator $\textrm{Na}_2\textrm{Zn}_2\textrm{Te}\textrm{O}_6$ \cite{Yao2020}, and subtracted from $C(T)$ to discuss the magnetic contribution \cite{SM}. The field dependence of magnetization $M(H)$ and the temperature dependence of magnetic susceptibility $\chi(T)$ were measured by using a commercial superconducting quantum interference device (SQUID) magnetometer.

\begin{figure}[t]
    \includegraphics[width=\linewidth]{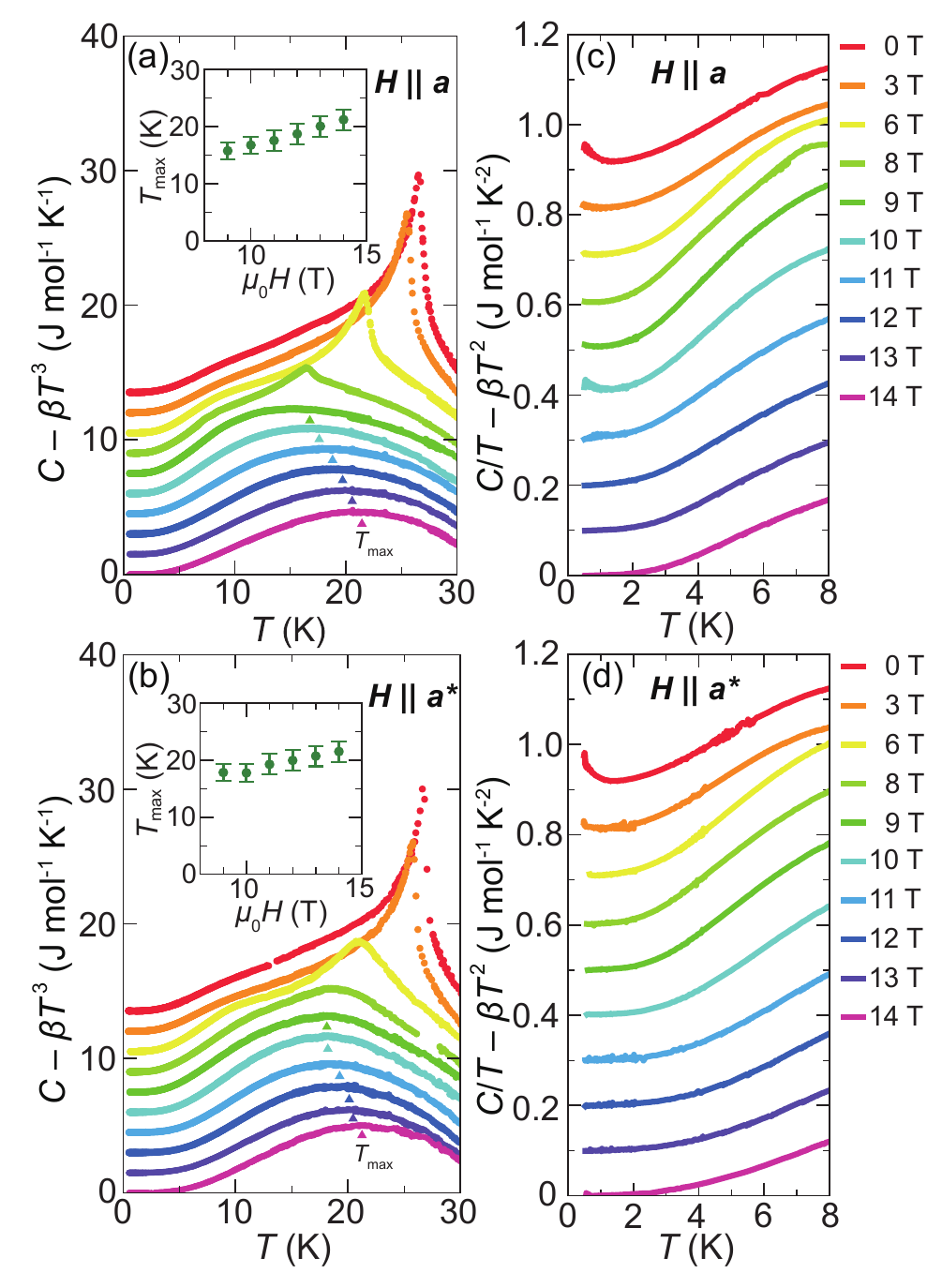}
    \caption{(a,b) Temperature dependence of specific heat $C$ with the phonon contribution $\beta T^3$ subtracted for \ha\ (a) and \has\ (b). 
    The data are vertically shifted for clarity. The insets show the field dependence of $T_{\rm max}$ at which $C(T)-\beta T^3$ has a broad peak (arrows in the main panel). (c,d) Temperature dependence of $C/T-\beta T^2$ for \ha\ (c) and \has\ (d). The data are vertically shifted for clarity.}
    \label{Fig2}
\end{figure}

Figures \ref{Fig2}(a) and (b) show the temperature dependence of $C-\beta T^3$ under several magnetic fields for \ha\ and \has, respectively. The magnetic phase transitions are determined by a sharp peak in $C(T)$ as well as the anomalies in magnetization and susceptibility measurements, from which we obtained the phase diagrams in Figs.\,\ref{Fig1}(b) and (c), which are consistent with the previous reports\,\cite{Yao2020,Lin2021,Hong2021,Takeda2022,Bera2023}. Above $\sim9$\,T, the transition anomaly in the specific heat becomes hardly discernible, and instead, $C(T)-\beta T^3$ has a broad peak at $T_{\rm max}$. The field dependence of $T_{\rm max}$ is shown in the insets of Figs.\,\ref{Fig2}(a) and (b), which shows an increasing trend with the field. These behaviors are qualitatively similar to the results in the high-field KSL state of \ru \cite{Tanaka2022}, in which the broad peak in $C(T)-\beta T^3$ has been attributed to a signature of the $Z_2$-flux excitations \cite{Tanaka2022,Motome2020}. This similarity in a relatively high-temperature regime suggests that Kitaev physics is at play in this energy scale, as discussed in the inelastic neutron scattering studies \cite{Songvilay2020,Kim2021}. We note that the values of $T_{\rm max}$, which can be used as a measure of $\Delta_{\rm flux}$ \cite{Tanaka2022}, are about two times higher than those in \ru. This may suggest a higher energy scale of Kitaev interaction in NCTO, as proposed theoretically in $3d$ honeycomb magnets. However, whether the KSL ground state is realized cannot be determined from such a high-temperature signature, and the presence of low-energy Majorana excitations must be examined.

\begin{figure}[t]
    \includegraphics[width=\linewidth]{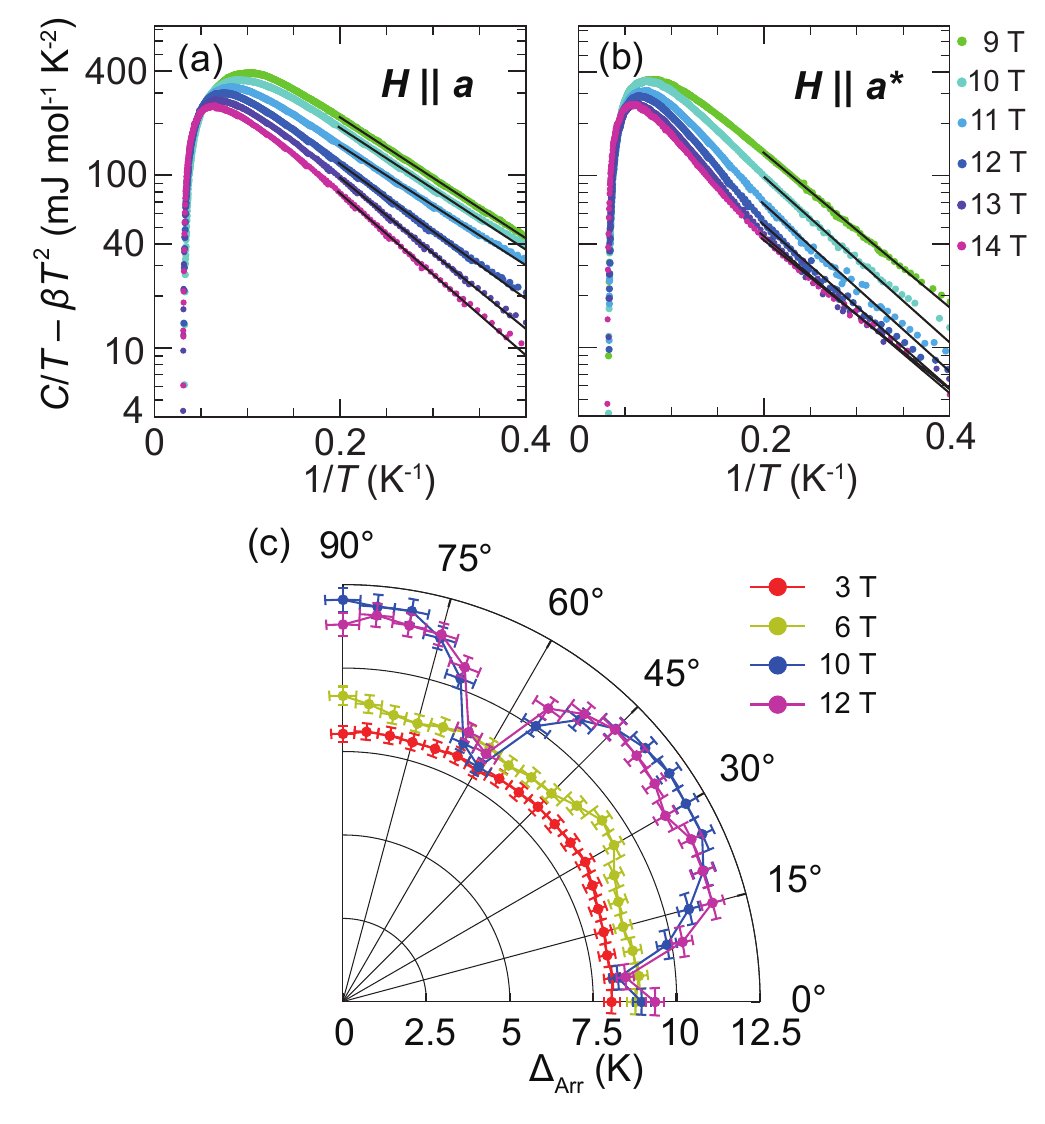}
    \caption{(a,b) Arrhenius plot of $C/T-\beta T^2$ for \ha\ (a) and \has (b). The black lines show the exponential fittings $e^{-\Delta_{\mathrm{arr}}/T}$ for a temperature range of 2.5-5\,K. The obtained gap values are shown in Figs.\,\ref{Fig1}(b) and (c). (c) Experimentally determined gap $\Delta_{\mathrm{arr}}$ as a function of field angle for several fields. The angle $0^\circ$ ($90^\circ$) corresponds to \ha\ (\has). }
    \label{Fig3}
\end{figure}

Next, we focus on the low-temperature behaviors of specific heat to discuss the low-energy excitations in the high-field state of NCTO. The low-$T$ behaviors of $C/T-\beta T^2$ are shown in Figs.\,\ref{Fig2}(c) and (d). The Schottky anomalies are prominent as an upturn at low temperatures below $\sim2$\,K and low fields. In \ha, an upturn can also be seen at 10\,T, which may be associated with a nontrivial magnetic structure near a critical field. To eliminate the possible effects of such anomalies, we focus on the data above 2.5\,K and the field angular dependence of specific heat, considering that the Schottky anomaly is expected to be insensitive to the field direction. 

As shown in Figs.\,\ref{Fig3} (a) and (b), we find that $C/T-\beta T^2$ shows exponential behaviors indicating gapped excitations for both \ha\ and \has. The excitation gap is estimated from the simple Arrhenius fitting $\sim e^{-\Delta_{\mathrm{arr}}/T}$ in a temperature range between 2.5 and 5\,K, where we find the fitting works reasonably well. 
We extend this analysis to different field angles and different field regimes, and the obtained results of the thermodynamic gap $\Delta_{\mathrm{arr}}(\phi, H)$ are shown in Fig.\,\ref{Fig3}(c) as well as in Figs.\,\ref{Fig1}(b) and (c).  
At low fields deep inside the AFM phase, the gap size is almost angle-independent, as shown in Fig.\,\ref{Fig3}(c). At higher fields above $\sim 6$\,T, the gap anisotropy starts to develop, and at 10\,T, we find that the gap depends strongly on the field angle. Remarkably, the high-field $\Delta_{\mathrm{arr}}(\phi)$ becomes the smallest for \ha\ and equivalent directions ($\phi=0^\circ$ and $60^\circ$), which is opposite to the largest gap expected from Eq.\,(\ref{deltam}) in the Kitaev model. Moreover, the gap should become zero for \has\ ($\phi=30^\circ$ and $90^\circ$) in the KSL, but we observe a gap significantly larger than that for \ha. We stress that such angle-dependent $\Delta_{\mathrm{arr}}(\phi)$ that emerges at high fields cannot be explained by Schottky anomalies, and thus it reflects the intrinsic properties of the low-energy excitations in the high-filed state of NCTO. The comparison of the field dependence of $\Delta_{\mathrm{arr}}$ for \ha\ and \has\ in Figs.\,\ref{Fig1}(b) and (c) shows that the anisotropy is most significant in the field range between $\sim 6$\,T and $\sim 13$\,T. It is noteworthy that this field range coincides with the range in which the finite $\kappa_{xy}$ is observed in NCTO \cite{Takeda2022}. 

\begin{figure}[t]
    \includegraphics[width=\linewidth]{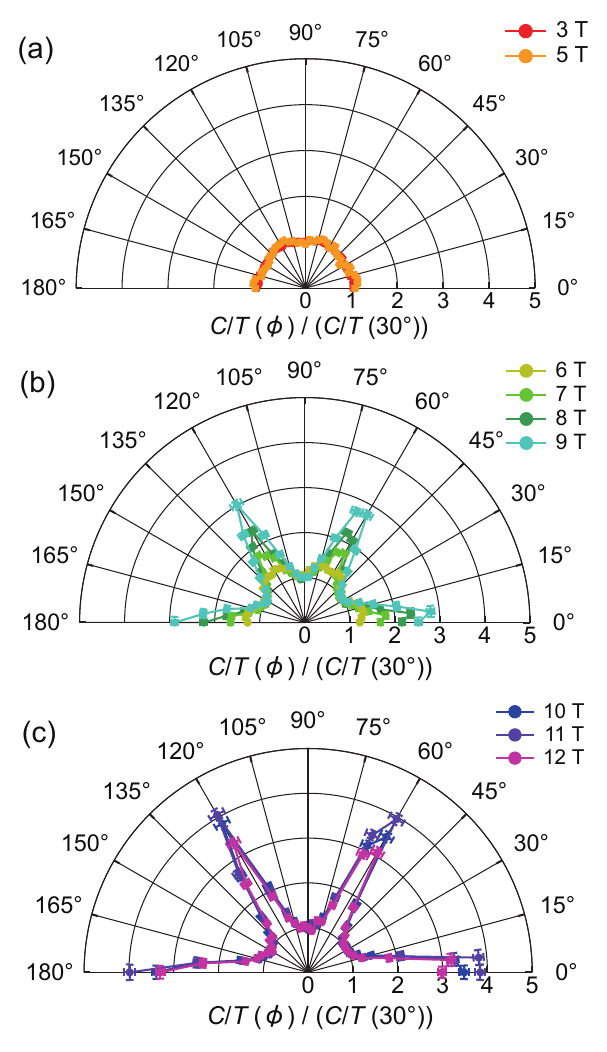}
    \caption{(a-c) Field-angle dependence of $C/T(\phi)$ at 2.5\,K normalized by the $\phi=30^{\circ}$ value for low fields (a), intermediate fields (b), and high fields above the critical fields (c). $\phi$ is the angle between the $\bm{a}$ axis and the field direction.}
    \label{Fig4}
\end{figure}

To compare our results in NCTO with the reported angle-resolved specific heat in the KSL phase of \ru\ \cite{Tanaka2022}, we plot the raw data of field-angle dependence of $C/T$ at a fixed temperature of 2.5\,K ($\ll T_{\rm max}$), which is low enough to discuss the intrinsic low-energy excitations. As shown in Fig.\,\ref{Fig4}(a), we observe isotropic $C/T$ at low fields, as found in the $\Delta_{\mathrm{arr}}(\phi)$ analysis. At higher fields, $C/T$ near \ha\ (and equivalent directions) starts to be enhanced, showing a peak-like structure with 6-fold rotational symmetry (Fig.\,\ref{Fig4}(b)), which is similar to that found in the AFM phase of \ru\ \cite{Tanaka2022}. Such a peak in \ru\ has been assigned to the spin-flop effect due to the field rotation. Above the critical fields ($\ge 10$\,T), the $\bm{a}$-axis peak becomes more prominent, which is completely different from the rounded minima near \ha\ found in the high-field spin-liquid phase of \ru \cite{SM}. 

The field-angle dependence of bulk low-energy excitations revealed by the specific heat is inconsistent with the Majorana quasiparticle excitations in the KSL. We note that the Majorana quasiparticles in the Kitaev model are the result of spin fractionalization, and thus may not be easily detected by spin probes. However, the Majorana fermions exhibit entropy and this is why the heat-related measurements, including thermal conductivity and specific heat, are powerful in KSL research. As such, our specific heat results provide thermodynamic evidence against the presence of a KSL in NCTO at least in the in-plane field range up to 14\,T. 

Finally, we discuss possible origins of the observed anisotropy of low-energy excitations, which may provide insights into the nature of the high-field state of NCTO. It has been recently reported that between the low-field AFM phase and the high-field disordered phase, a canted antiferromagnetic state exists up to 10\,T in NCTO \cite{Bera2023}. In such a case, calculations using the Kitaev-Heisenberg model have shown that the magnon excitation gap can show strong field-angle dependence with sharp minima along the $\bm{a}$ axis \cite{Koyama2021}, which is consistent with our observation. These calculations also show that the finite thermal Hall conductivity can appear due to the topological nature of the spin-wave dispersions. 
Although no clear magnetic order is seen above 10\,T, the strong anisotropy in $C/T(\phi)$ suggests that similar excitations of topological magnons persist at high fields. This is supported by the thermal Hall conductivity measurements in NCTO, in which even above the critical fields, finite $\kappa_{xy}$ is observed up to $\sim 14$\,T \cite{Takeda2022}. 

In the high-field state of \ru, the sample dependence leads to debates on the origin of the in-plane thermal Hall effect. As a possible scenario, it has been suggested that topological magnons may be responsible for the observed $\kappa_{xy}$ \cite{Czajka2023}. Indeed, theoretical studies have shown that such a scenario is possible in a partially polarized state above the critical field \cite{Zhang2021}. In \ru, this magnon origin is not compatible with the field-angle dependence of the excitation gap, which is consistent with the KSL \cite{Tanaka2022,Imamura2024}. In contrast, the present angle dependence of fully gapped excitations in the high-field state of NCTO is more consistent with such a magnon scenario than a KSL. 
The reason why these two systems have different properties at high fields deserves further investigation. Still a recent theory pointed out that the relationship between the Kitaev and Heisenberg interactions in cobaltates may be sensitively determined by the inter- and intra-orbital exchange paths \cite{Liu2023}, implying that small differences in crystal structures can lead to different ground states. 

In summary, our field-angle-resolved specific heat measurements revealed the absence of the gapless excitation for \has, indicating that NCTO does not exhibit a KSL under in-plane magnetic fields. Instead, finite gaps were observed across all magnetic phases, and the anisotropy in the gap size suggests that these gaps at high fields originate from topological magnon excitations. These findings not only provide critical insights into the low-energy excitation spectrum of NCTO, but also demonstrate that the field-angle dependence is important in the Kitaev physics research.

We thank E.-G. Moon, J.-G. Park, H. Takeda, and N. Trivedi for fruitful discussions. This work was supported by Grant-in-Aid for Scientific Research (KAKENHI) (No. JP23H00089, No. JP24H00937, No. JP24H01640, No. JP23K26522, No. JP21KK0242), Grant-in-Aid for Scientific Research on innovative areas ``Quantum Liquid Crystals'' (No. JP19H05824), Grant-in-Aid for Scientific Research for Transformative Research Areas (A) ``Condensed Conjugation'' (No. JP20H05869) from Japan Society for the Promotion of Science (JSPS), and CREST (No. JPMJCR19T5) from Japan Science and Technology (JST).

\bibliography{NCTO}

\end{document}